# IMPLEMENTING AND EVALUATING A WIRELESS BODY SENSOR SYSTEM FOR AUTOMATED PHYSIOLOGICAL DATA ACQUISITION AT HOME

Chao Chen and Carlos Pomalaza-Ráez

Department of Engineering, Indiana University – Purdue University
Fort Wayne, Indiana, USA
`chen@engr.ipfw.edu`
`carlos-pomalaza-raez@purdue.edu`

## ABSTRACT

*Advances in embedded devices and wireless sensor networks have resulted in new and inexpensive health care solutions. This paper describes the implementation and the evaluation of a wireless body sensor system that monitors human physiological data at home. Specifically, a waist-mounted triaxial accelerometer unit is used to record human movements. Sampled data are transmitted using an IEEE 802.15.4 wireless transceiver to a data logger unit. The wearable sensor unit is light, small, and consumes low energy, which allows for inexpensive and unobtrusive monitoring during normal daily activities at home. The acceleration measurement tests show that it is possible to classify different human motion through the acceleration reading. The 802.15.4 wireless signal quality is also tested in typical home scenarios. Measurement results show that even with interference from nearby IEEE 802.11 signals and microwave ovens, the data delivery performance is satisfactory and can be improved by selecting an appropriate channel. Moreover, we found that the wireless signal can be attenuated by housing materials, home appliances, and even plants. Therefore, the deployment of wireless body sensor systems at home needs to take all these factors into consideration.*

## 1. INTRODUCTION

With the advances in embedded microcontrollers, inexpensive miniature sensors, and wireless networking technologies, there has been a growing interest in using wireless sensor networks in medical applications. For example, wireless sensor networks can replace expensive and cumbersome wired devices for pre-hospital and ambulatory emergency care when real-time and continuous monitoring of vital signs is needed. Moreover, body sensor networks can be formed by placing low-power wireless devices on or around the body, enabling long-term monitoring of physiological data. For elderly patients and people with chronic diseases, an in-house wireless sensor network allows convenient collection of medical data while they are staying at home, thus reducing the burden of hospital stay. The collected data can be passed onto the Internet through a PDA, a cell-phone, or a home computer. The caregivers thus have remote access to the patient's health status, facilitating long-term rehabilitation and early detection of certain physical diseases. If there are abnormal changes in the patient status, caregivers can be notified in a timely manner, and immediate treatment can be provided.

Applying wireless sensor networks for health monitoring faces several key challenges.

- To achieve non-obtrusive continuous health monitoring, wireless sensors used in medical applications need to be light, small, and wearable. For safety considerations, the electromagnetic signal needs to be low enough to limit human exposure to radio frequency (RF) radiation [1][2].

- Power and memory are limited for on-board signal processing and computation since wearable sensors are battery-operated. Power is also needed for transferring raw data. Therefore, a careful trade-off design between communication and on-board data processing is crucial for an optimal system design. Moreover, low-duty cycle operations need to be explored to extend the battery life.

- Medical monitoring applications require high reliability in data collection. The RF signal inevitably faces absorption and interference from medical equipments in a hospital environment [3]. Even in a home environment, the interference cannot be ignored because of the growing popularity of home devices with wireless interfaces.

- System security is another important issue for medical applications [4]. Wireless medical sensors must meet privacy requirements mandated by the law for all medical devices and must guarantee data integrity. Such requirements can be met by authentication and data encryption, which increase the computation and energy consumption of resource-constrained sensors.

In this work, we implement and evaluate a wearable sensor system for monitoring a patient's physical activities at home. As a demonstrative example, we select accelerometers to measure human movements. Acceleration measurements are wirelessly transmitted through an IEEE 802.15.4 compliant RF transceiver and recorded for post analysis. Specifically, our work makes the following contributions:

- We have built the hardware platform of a wireless body sensor system that consists of a wearable sensor unit and a data logger unit. The wearable sensor unit is light, small, and energy-efficient. We also designed a workflow of the accelerometer to further reduce the power consumption without affecting measurement accuracy. The wearable sensor unit can be attached to the body and operate for days without changing battery. The data logger unit can store days of motion data continuously using a memory card. The system prototype allows for inexpensive and unobtrusive monitoring during normal daily activities at home.

- Simple analysis of the recorded data shows that different types of human movements generate different acceleration patterns. Therefore, it is possible to classify different human motion through the analysis of the recorded acceleration data. Moreover, abrupt movements can be detected from abnormal acceleration readings. We also suggest some simple methods to detect abnormal activities by observing the acceleration reading.

- To evaluate the feasibility of implementing an IEEE 802.15.4-based wireless medical sensor system at home, we have performed a series of measurement tests at both an apartment scenario and a single-house scenario. The impact of wireless interference and signal attenuation on the performance of 802.15.4 transmission is evaluated. Our measurement results show that 802.15.4 transmission is affected by nearby wireless signals that occupy the same frequency band (*e.g.*, IEEE 802.11 signals and microwave ovens). To reduce such impact, the 802.15.4 transceivers can adaptively select an appropriate operating channel with less interference. We also find that housing materials, home appliances, and even plants can attenuate the wireless signal at different scales. Therefore, the deployment of wireless body sensor systems for smart home health care needs to take these factors into account.

The remainder of this paper is organized as follows. Section 2 gives the overview and the selection of the key components of the wireless body sensor system. Section 3 describes the architecture and the design of the system prototype. Section 4 presents the measurement tests of human physical activities, including slow movements and abrupt movements. Section 5 gives the interference tests and signal attenuation tests of 802.15.4 signal transmission at home. Section 6 concludes this paper.

## 2. SYSTEM OVERVIEW

In this section, we give an overview of our wireless body sensor system. The selection of the sensor technology and the wireless technology involved in the system design is discussed. Since this system is planned for use in a private environment and only consists of a couple of wireless nodes with a short communication range, system security is not a major concern.

Our designed wireless body sensor system consists of a *wearable sensor unit* and a *data logger unit*, as shown in Fig. 1. The wearable sensor unit acquires a person's physiological signals. The measured data are sampled via an analog-to-digital converter (ADC) and fed into a microcontroller. The sampled data are then transmitted wirelessly to the data logger unit, which is fixed at a location close to a grid power source. The measured data are downloaded to a computer through a cable connection and analyzed using software programs. The collected data can also be passed onto a public network (*e.g.*, Internet or cellular networks) so that the caregivers have remote access to the patient's health status.

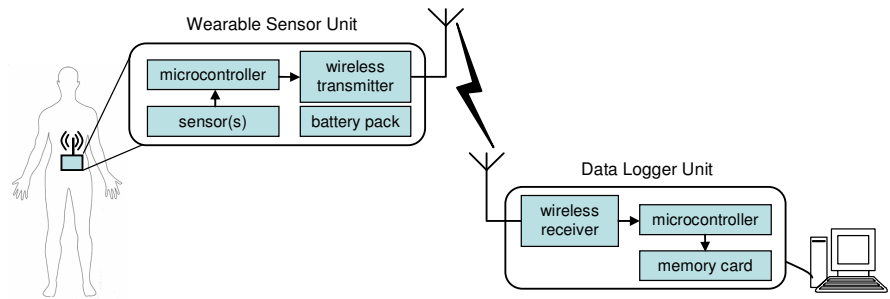

Figure 1   A wireless body sensor system for physiological data acquisition

### 2.1. Selection of Sensor Technology

Different types of sensors can be embedded into the wearable sensor unit for medical use. Examples include an electrocardiogram (ECG) sensor to monitor heart activity, an electromyogram (EMG) sensor to monitor muscle activity, an electroencephalogram (EEG) sensor to monitor brain electrical activity, and a blood pressure sensor to monitor blood pressure. For demonstration purpose, the wearable sensor unit in our system monitors human movements only.

Inertial measurement components sense either translational acceleration or angular rate. The advances in micro-electromechanical systems (MEMS) and other micro-fabrication techniques have greatly reduced the cost and the size of these devices, and they can be easily embedded into wireless and mobile platforms. *Gyroscopes* and *accelerometers* are two common inertial sensors that can be used to capture human motion continuously. An accelerometer measures the acceleration relative to freefall, including the gravitational acceleration *g*. Movement in three dimensions can be determined using accelerometer measurements over three orthogonal axes. A gyroscope measures angular velocity. If the initial angle is known, integration over time provides the change in angle. Compared with accelerometers, gyroscopes have drawbacks such as higher power consumption, higher cost, possibility of drift, and sensitivity to shock. In this work, therefore, we choose accelerometers in the wearable sensor unit.

There are several common types of accelerometers [5]: *piezoelectric* accelerometers that exploit the piezoelectric effect of certain quartz crystals that produce an electric charge in response to applied mechanical stress; *piezoresistive* accelerometers by measuring the resistance of a material that changes under mechanical stress; capacitive accelerometers where the change in capacitance is proportional to acceleration; and *hall effect* accelerometers that operate by converting motion to an electrical signal through sensing of changing magnetic fields. The capacitive accelerometer was chosen to measure human motion in our work because of its low price, small size, capability of continuous measurement, and ease of integration.

## 2.2. Selection of Wireless Technology

Our design is targeted for indoor use such as in a house and a nursing home environment. Therefore, a short-distance wireless communication system is appropriate. Two types of wireless communication standards are suitable for this application: *IEEE 802.15.1 (Bluetooth)* [6] for medium rate wireless personal area networks (WPAN) and *IEEE 802.15.4 (ZigBee)* [7] for low rate WPANs, both operating in the 2.4 GHz unlicensed industrial, scientific, and medical (ISM) frequency band.

IEEE 802.15.1 is adapted from Bluetooth, which is a telecommunications industry specification for short-range RF-based connectivity for portable devices. Bluetooth is designed for small and low cost devices with low power consumption. Since Bluetooth is geared towards handling voice, images, and file transfer, it has a data transfer rate on the order of 1 Mbps with a relatively complex protocol. The operational range for Bluetooth is around 10 meters. With amplifier antennas its range can be boosted to 100 meters, but with higher power consumption.

IEEE 802.15.4 specifies physical layer and medium access control for WPANs which focus on low-cost, low-speed ubiquitous communication between devices. It is designed for systems that need a battery life as long as several months to several years but do not require a data transfer rate as high as those enabled by Bluetooth. The 802.15.4 compliant devices have a transmission range between 10 and 75 meters and a data transfer rate of 250 kbps (if operating at 2.4 GHz frequency band). 802.15.4 supports a basic master-slave configuration suited to static star networks of many infrequently used devices that talk via small data packets. Compared with Bluetooth, 802.15.4 is more power-efficient because of its small packet size, reduced transceiver duty cycle, reduced complexity, and strict power management mechanisms such as power-down and sleep modes.

Both IEEE 802.15.1 and IEEE 802.15.4 have been used as the wireless communication protocol in different wearable sensing platforms for measuring and recognize human activities, *e.g.*, [8][9]. After comparing the two choices of applicable wireless standards, it is clear that IEEE 802.15.4 is the better choice for our proposed design that requires short range wireless communication between low-cost, low-power, and battery-operated devices for monitoring purposes.

## 3. SYSTEM IMPLEMENTATION

In this section, we describe the detailed design of the wireless body sensor system, including hardware design, accelerometer workflow, and system power, size, weight, and cost analysis.

## 3.1. Wearable Sensor Unit

The Freescale MMA7260QT triaxial capacitive micro-machined accelerometer [10] is selected as the measurement device on the wearable sensor unit. This device has two surface micro-machined capacitive sensing cells (g-cells) where the center beam moves with acceleration. The accelerometer also has a signal conditioning ASIC that uses switched capacitor techniques to measure the g-cell capacitors and extract the acceleration data. The ASIC also conditions and filters the signal, providing a high level output voltage that is proportional to acceleration. MMA7260QT has four different measurement ranges ($\pm1.5g$, $\pm2.0g$, $\pm4.0g$, and $\pm6.0g$) that can be dynamically set by two *g*-Select input pins. Each range provides different measurement sensitivity levels (800mV/*g*, 600mV/*g*, 300mV/*g*, and 200mV/*g*). The accelerometer continuously records human movements in terms of accelerations in all three axes. The lower ranges (*e.g.*, ±1.5*g*) are used primarily for accurate measurements of small motions, whereas the higher ranges (*e.g.*, ±6.0*g*) are used primarily for measurements of vibration and impact generated during activities such as running and falling. The MMA7260QT accelerometer has a low power consumption with 2.2V~3.6V voltage supply and 500 µA current flow at the normal condition. It can also be set to a low-current inactive mode (*i.e.,* sleep mode) of only 3 µA operation current through a SLEEP pin, which further conserves power.

The measured data are read by a Silicon Labs C8051F353 microcontroller with a built-in 16-bit ADC [11]. The microcontroller does simple processing on the data and sets the working mode of the accelerometer accordingly. The processed data are fed to a wireless transceiver and sent to the data logger unit.

We designed the workflow of the accelerometer, as shown in Fig. 2. Initially the accelerometer is set in sleep mode and wakes up periodically at a low frequency (*e.g.,* every 1 sec), taking a sample with a low measurement range (*e.g.,* ±1.5*g*) and high measurement sensitivity. This allows minimum power consumption when a person is in rest mode. Once a measurement reading exceeds a certain threshold value, which means that this person is now in active motion, continuous measurement will be taken with the sleep mode disabled. The sampling rate is set by the microcontroller to be in the range of 10-100 Hz, as it has been found that the frequency of human induced activity ranges from 1 to 18 Hz [12]. The accelerometer's measurement range is set adaptively to the actual measurement. For example, if the current range is ±2*g* with a sensitivity of 600mV/*g* and the measurement is below -2*g* or above 2*g*, then the next measurement will be taken with a higher range of ±4*g* with a lower sensitivity of 300mV/*g*; if the measured acceleration is between -1.5*g* and 1.5*g*, then the next measurement will be taken with a lower range of ±1.5*g* with a higher sensitivity of 800mV/*g*. The purpose is to allow for the highest sensitivity level possible while getting the closest reading. Furthermore, the measurement range is adjusted independently in different axes. If the acceleration measurements show that the movements have stayed at a low level continuously for a while (*e.g.,* for a duration of 5 min), the accelerometer will go back to the sleep mode and wake up periodically, checking for possible active motion. In summary, this workflow enables automated sampling of acceleration data, saves power during inactive periods, and gets timely data measurements during active periods.

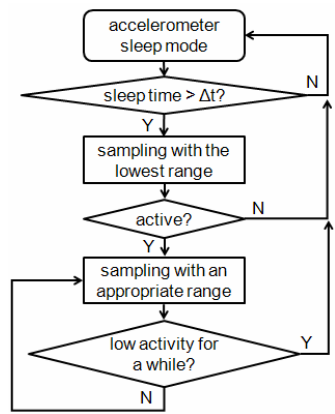

Figure 2   Workflow of the accelerometer

The XBee® 802.15.4 radio modem from MaxStream [13] is chosen as the wireless transceiver. It can operate under a transparent mode with a simple connection with a microcontroller. With a chip antenna, it operates up to 30 meters indoor. The transmission range can be further increased to 90 meters by using a whip antenna. The XBee module has a low transmit power (≤1 mW) and a high receiver sensitivity of -92 dBm.

### 3.2. Data Logger Unit

The front-end of the data logger unit is a wireless XBee transceiver. Upon receiving the measurement data from the wireless interface, the XBee transceiver forwards the data directly to the microcontroller for processing. We chose C8051F344 [14] from Silicon Labs as the microcontroller in the data logger unit. It has a convenient serial interface with a flash memory card. The processed data then serve as the basis for the calculations and software development involved in the characterization of movements.

### 3.3. Prototype Design and Analysis

Fig. 3(a) shows the wearable sensor unit prototype. Besides the accelerometer, the microcontroller, and the XBee transceiver, the other components are: a switch for power on/off, an interface port to program the microcontroller, some peripheral circuitry, and a battery power. We select a 3.7 V Li-Ion rechargeable battery pack with 6600 mAh of energy as the power source. In the wearable sensor unit, most power is consumed by the wireless transceiver. In the worst case, if the XBee RF modem is continuously transmitting, the wearable unit can be used for more than 120 hours without recharging. If the transceiver is active for 10% of the time and in standby mode for 90% of the time, the wearable sensor unit can be operated for more than 600 hours (~25 days) without changing battery. Therefore, adding some pre-processing in the microcontroller and dynamically changing the sampling rate based the human activity level can reduce the frequency and amount of data transmitted over the air, which in turn saves energy.

The prototype of the data logger unit is given in Fig. 3(b). Besides the XBee transceiver and the microcontroller, the following components are also included: a secure digital memory card with the interface board, an RS232 port to download the contents of the memory card to a PC, a switch to control the downloading of card data, an interface port to program the microcontroller, a power jack, and some peripheral circuitry.

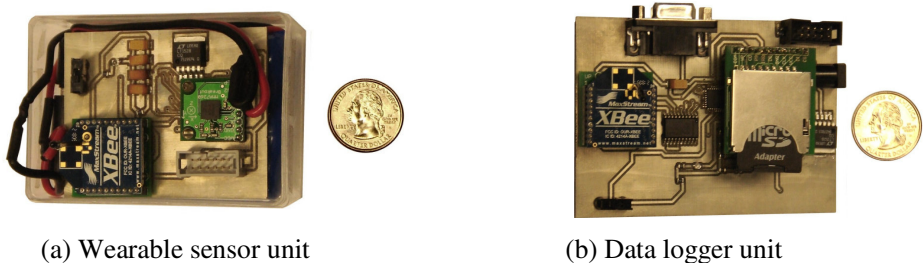

(a) Wearable sensor unit           (b) Data logger unit

Figure 3  System prototype

The power, size, and weight analysis for the major components in the wearable sensor unit is listed in Table 1. The size of the whole sensor unit prototype (including the casing) is limited to 75×60×30 mm, and the weight is less than 270 g. The whole unit can be wearable without much disturbance to the person. The size and the weight can be further reduced by selecting a battery pack that is lighter and smaller. In our current prototype design, the Li-Ion battery pack accounts for ~95% of the total weight; it was selected mainly because of its low cost and the convenience of testing and recharging. If smaller size and less weight are preferred, other types of batteries can be used. For example, the LIR3048 Li-ion rechargeable button cell battery has a small size of 30×30×4.8 mm and a light weight of 12 g for each cell; but its low capacity (230 mAh) only lasts for around 4 hours for continuous operation and about 21 hours for 10% operation (longer operation life can be provided by using a few cells together). Since the data logger unit is connected to a fixed power source, the power consumption, size, and weight will not be an issue.

Table 1   Power, size, and weight analysis of the wearable sensor unit

| Component | Typical current (mA) | Size (mm) | Weight (g) |
|---|---|---|---|
| MMA7260QT accelerometer | 0.5 | 6×6×1.45 | 1.5 |
| C8051F353 microcontroller | 5.8 | 9×9×1.6 | 1.5 |
| XBee 802.15.4 transmitter | 45 | 25×28×2.8 | 4 |
| Li-Ion 18650 battery pack | - | 69×54×18 | 255 |

The cost to build the overall system prototype (excluding the PC) is around 150 dollars, which makes it feasible for home use. Moreover, a single data logger unit can be used along with several wearable sensor units because of the built-in capability of the XBee RF module to form a star network. This allows multiple users to share the same data logger unit, which is especially convenient in a nursing home scenario.

## 4. ACCELERATION MEASUREMENT TESTS

Experimental research has shown that normal daily activities can be recorded by attaching accelerometers to the human body with minimum discomfort for the subject. Different body locations for sensor attachment have been proposed for various reasons such as the convenience of measurement and targeted medical applications. For example, in [12], triaxial accelerometers were placed on the back waistline to study the relationship between metabolic energy expenditure and different physical activities. A kinematic sensor consists of two accelerometers and a gyroscope was attached to a subject's chest as part of an ambulatory measurement system [15]. A sensor platform with two accelerometer components was placed on an ankle for step recognition [16]. A wearable wireless accelerometer was attached to the wrist to record early morning bathroom activities for assisting patients with cognitive impairments due to traumatic brain injury [17]. Since accelerometers placed at the waist have been used to resolve resting states, such as sitting, standing, and lying, and activities including walking, falling, and transition between resting states [18][19], we tested the accelerometer measurement by attaching the wearable sensor unit to a person's waist, like shown in Fig. 4. The sampling frequency is 60 Hz for each axis.

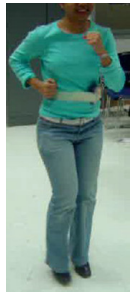

Figure 4   Attachment of the wearable sensor unit

Fig. 5 shows the acceleration measurements over three axes during three activity tests with slow movements: sit-stand transitions, left-right rotations, and walking at a slow pace. $a_x$, $a_y$, and $a_z$ refer to the frontal, side, and vertical accelerations, respectively. Because of the Earth's gravity, a positive $1g$ output is always present in the vertical measurement $a_z$, if the person is in a vertical position. In these testing cases, since the movement is mainly in the frontal and side axes, the acceleration over $z$-axis does not depict an obvious pattern. The total acceleration $a$ ($= \sqrt{a_x^2 + a_y^2 + a_z^2}$) for each activity type is plotted in a separate figure. For these slow movements, the total acceleration always lies in the range of [0.9$g$, 1.3$g$].

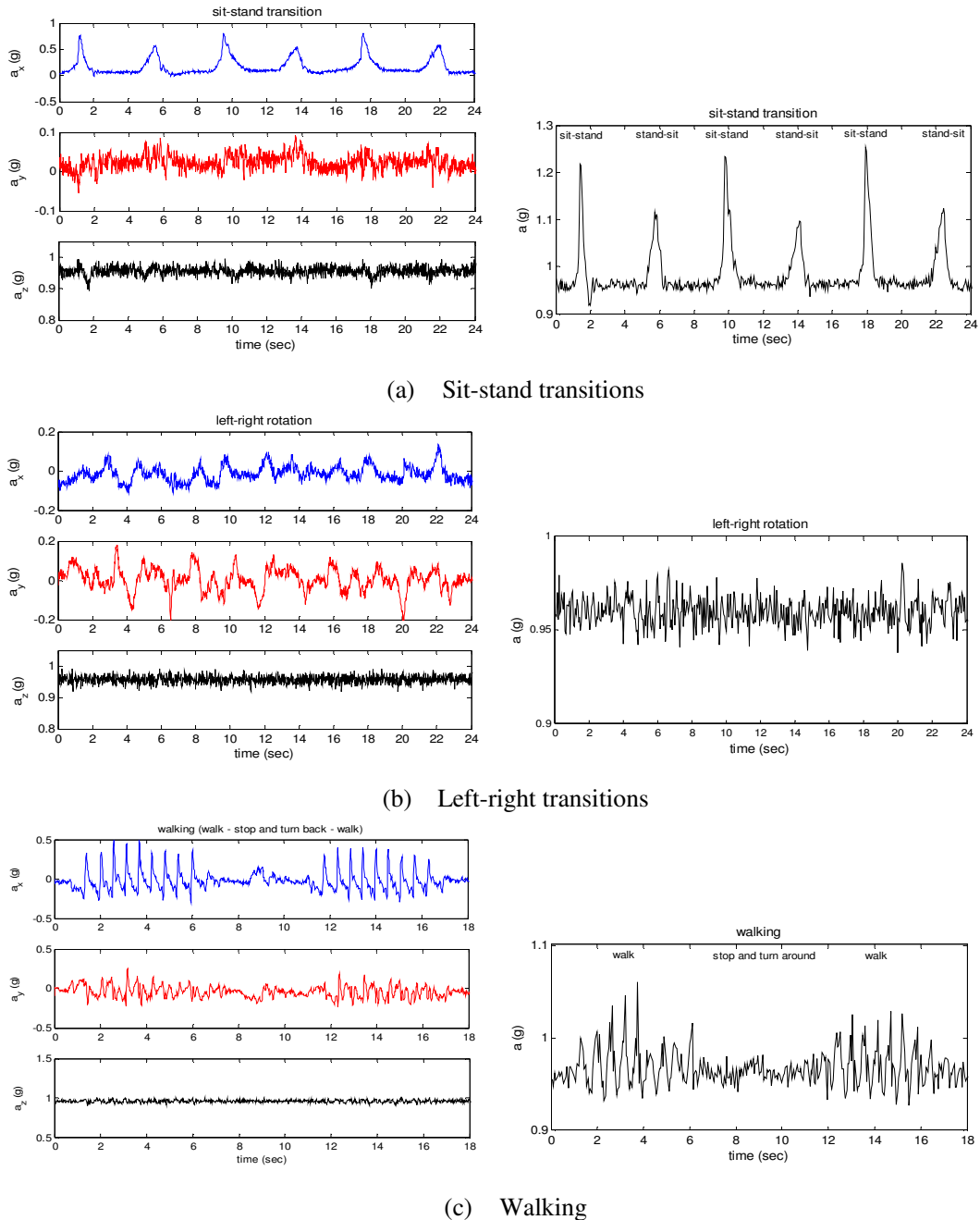

(a) Sit-stand transitions

(b) Left-right transitions

(c) Walking

Figure 5  Acceleration measurements of movement tests with slow-mode activitie

Another set of movement tests with faster activities (running, jumping, and falling) were conducted and the accelerations over three axes are recorded in Fig. 6. In these tests, the signals corresponding to the *z*-axis show greater variation than for the slow-mode cases in Fig. 6. In the jumping and falling tests with more abrupt movements, the maximum range of acceleration change over *z*-axis is above 2*g*. When the total acceleration *a* is considered, we can see abnormal values that are smaller than 0.9*g* or greater than 1.3*g*. The measurement results suggest two simple methods to detect abnormal activities by observing the acceleration reading: i) monitoring the acceleration change over each axis, and ii) checking the total acceleration to see if it is above a high limit or below a low threshold.

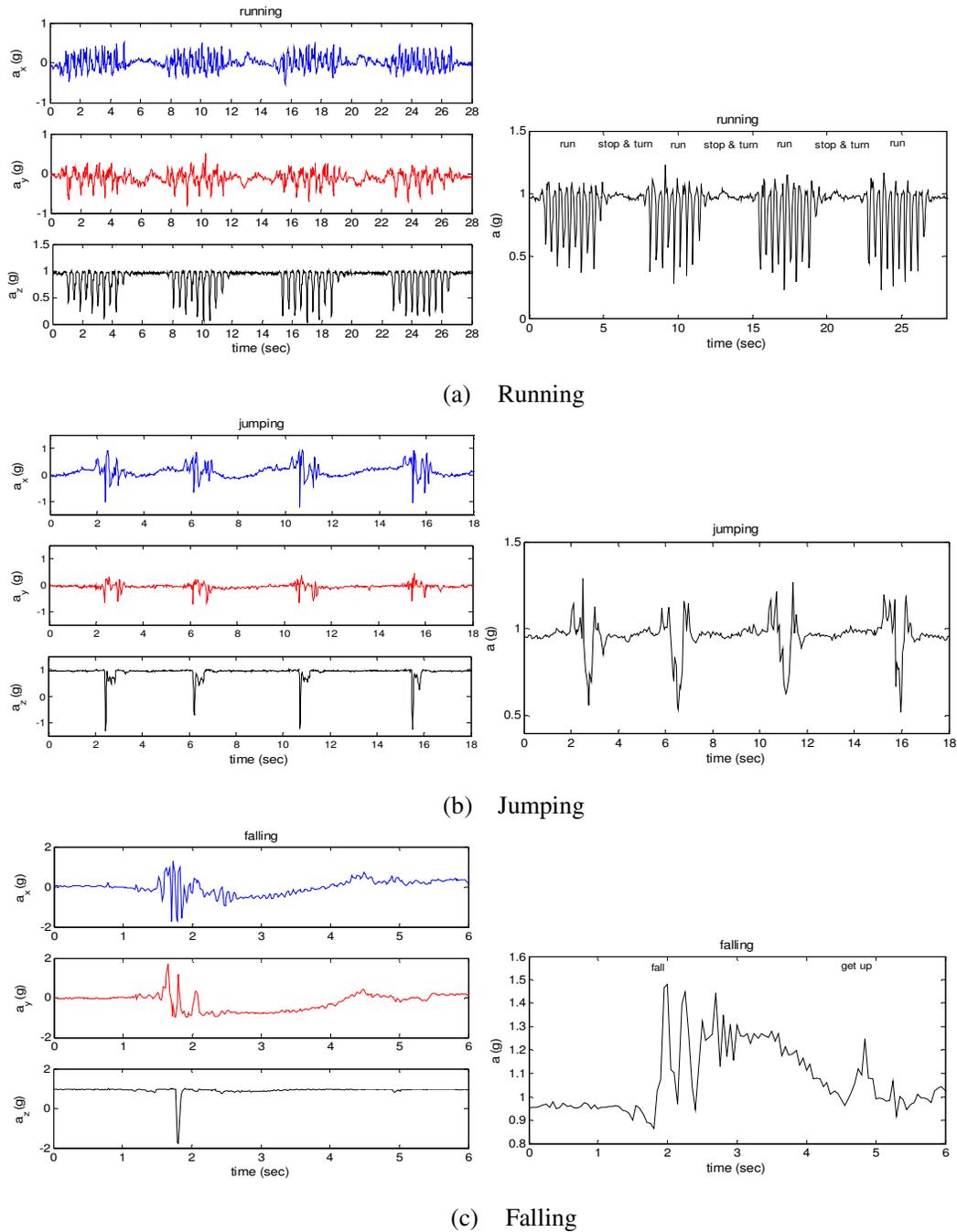

(a) Running

(b) Jumping

(c) Falling

Figure 6    Acceleration measurements of movement tests with fast-mode activities

The above measurement results show that different types of human motion generate different patterns in the acceleration reading. Therefore, it is possible to classify different human motion through the analysis of the recorded acceleration data and warn against abnormal activities such as falling. Moreover, once an abnormal activity is detected through the acceleration reading, the health status can be verified with inputs from other sources, *e.g.,* by checking the heart activity from an ECG sensor, or activating a camera or a motion sensor at home.

## 5. WIRELESS SIGNAL QUALITY TESTS

Medical monitoring applications require high reliability in data collection. To test the feasibility of using IEEE 802.15.4 standard for health care applications at home, we performed several measurement tests at typical home scenarios in the United States. Specifically, we focus on the following two major challenges.

- *Interference from other wireless systems:* 802.15.4 operates in the 2.4 GHz license-free ISM band, which is shared by several other wireless standards, such as the IEEE 802.11 WLAN. With the growing popularity of home devices with WLAN interfaces, wireless monitoring systems at home using 802.15.4 standard will inevitably face interference from these devices. Moreover, most household microwave ovens operate on the 2.45 GHz frequency. The interference from microwave ovens at home cannot be ignored either.

- *Signal attenuation:* To reduce human exposure to RF signal, the 802.15.4 transmitter needs to operate on a low power. For security, low power can also reduce the possibility of eavesdropping from outside. However, the receiver requires a certain signal quality level to correctly interpret the data (*e.g.,* SNR $\geq$ -92 dBm for the XBee module in our system). Besides the transmitting distance, there are other factors in a typical home scenario that would attenuate the signal quality. Example sources of attenuation include house appliances, furniture, building materials, *etc*.

We have performed a series of experimental tests to evaluate the performance of 802.15.4 transmission at home. Specifically, we used an XBee 802.15.4 development kit that comes with two XBee RF modules, as shown in Fig. 7, and a software program. One module (*i.e.,* the "base" module) is mounted to a USB board that connects to a PC. The other module (*i.e.*, the "remote" module) functions as a repeater by looping data back for retransmission. The operating channel and transmit power of the RF modules can be easily adjusted through the software program. In every test run, the base module sends out 1000 identical data messages, each with 32 characters in length. The remote module receives and retransmits data back. The timeout value for data reception is set as 100 msec.

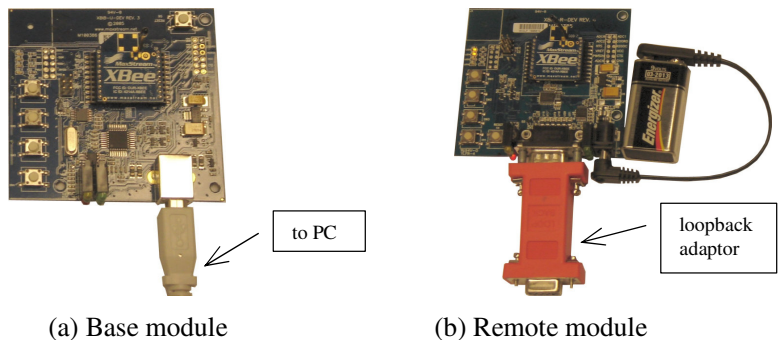

(a) Base module          (b) Remote module
Figure 7    The wireless signal measurement test-bed

### 5.1. Wireless Interference Tests

#### 5.1.1. Co-existence with 802.11

Compared with WPANs, WLANs have a longer transmission range (around 100 meters) and higher power consumption. IEEE 802.11 standard specifies medium access layer and physical layer mechanisms for WLANs [20]. Among the set of 802.11 standards, 802.11b and 802.11g have become extremely popular since their introduction because of their lower cost and longer range compared with 802.11a. Therefore, we focus on the interference from 802.11b and 802.11g, and from this point on in this paper 802.11 refers to 802.11b/g. Both 802.11b and 802.11g operate on the 2.4 GHz frequency band with the same channel allocation. The overlapping frequency allocations of 802.11b/g and 802.15.4 in the United States are shown in Fig. 8. Most WLAN wireless routers operate on one of the non-overlapping channels (*i.e.*, channels 1, 6, or 11) with a 25 MHz bandwidth. 802.15.4 devices in the 2.4 GHz band have 16 non-overlapping channels to choose from, each with a 5 MHz bandwidth. The output power of 802.11 devices is usually at 15 dBm or above, which is much higher than the power of 802.15.4 devices ($\leq$ 0 dBm).

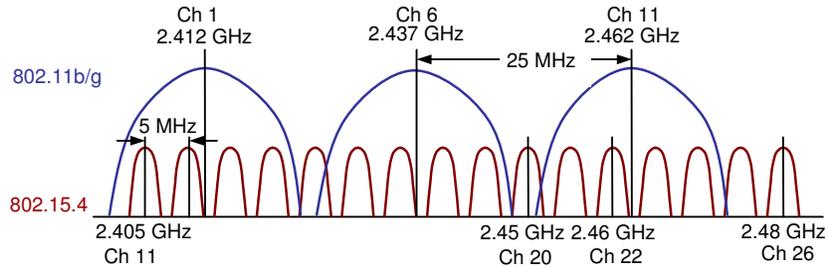

Figure 8    802.11b/g and 802.15.4 channels in the 2.4 GHz band

There have been some experimental tests to evaluate the impact of IEEE 802.11 on the performance of IEEE 802.15.4, for example, [21][22][23]. Their work indicates that 802.11 has a negative effect on the performance of 802.15.4 transmission, and there should leave some offset in their operational frequencies for a satisfactory performance of 802.15.4. However, their work focuses on the worst case by deliberately putting 802.11 and 802.15.4 devices right next to each other. On the other hand, our measurement tests were done in residential homes, where the interference came from WLANs inside home and from surrounding neighbours. We believe that our test environment is closer to the proposed smart home health care scenario.

The first set of tests was performed in a one-bedroom apartment. The longest distance between two points in the apartment is about 10 meters, which is within the transmission range of the XBee module. The two modules should be able to receive each other's data successfully if no interference is present. In this test scenario, neighbors live close in the apartment building. During our measurement tests, eight 802.11g networks were consistently present. Among them, three were with good to very good signal strength (Two were operating on channel 1, and one was on channel 11). The ones with fair but recognizable signals were scattered in channels 1, 6, and 11. In addition, a total of nine other 802.11g networks also appeared during the tests.

We first set our XBee module to work on channel 12 with a center frequency of 2.41 GHz. This is a worst case that the offset to the strongest 802.11 channel (*i.e.*, channel 1 in our case) is only 2 MHz. We tested the data success ratio under different power levels, and the results are listed in Table 2. The result for each test case is an average of 10 runs. When the XBee modules are transmitting at a lower power of -10 dBm, the average data success ratio is slightly lower and the standard deviation is higher. Next, we let XBee modules operate on channel 20 with a center frequency of 2.45 GHz, which is in the middle between 802.11 channel 6 and channel 11. In this case, the data success ratio is always 100% even when the XBee modules are transmitting at a low power level of -10 dBm.

Table 2    802.11 interference test in the apartment scenario

| Power level of XBee modules | 802.15.4 channel | Data success ratio | |
|---|---|---|---|
| | | mean | std. |
| 0 dBm (1 mW) | 12 | 99.36% | 0.39% |
| -10 dBm (100 μW) | 12 | 99.21% | 0.75% |
| -10 dBm (100 μW) | 20 | 100% | 0% |

Table 3   802.11 interference test in the single-house scenario

| Power level of XBee modules | 802.15.4 channel | Data success ratio | |
|---|---|---|---|
| | | Mean | std. |
| 0 dBm (1 mW) | 22 | 99.89% | 0.15% |
| -10 dBm (100 μW) | 22 | 99.66% | 0.23% |
| -10 dBm (100 μW) | 20 | 99.91% | 0.08% |

The same test was performed in a single house with two floors and a basement. During the measurement tests, one 802.11g WLAN using channel 11 was active inside the house, and three other different 802.11g neighborhood networks were present. The distance from the neighbors is at least 12 meters. Therefore, the wireless interference from the neighbors is much less compared to the WLAN inside the house. In our tests, the base module was in the living room on the first floor and the remote module was in a room on the second floor, which is roughly 6 meters away. To test the performance of XBee module in the worst case, 802.15.4 channel 22 was selected with a center frequency of 2.46 GHz. Moreover, we also tested a better case where the XBee modules are operating at 802.15.4 channel 20. The data success ratio of the tests in this scenario is shown in Table 3, where the result for each test case is an average of 10 runs.

The results show that 802.15.4 devices did experience interference from 802.11 signals. If the user application requires a high reliability, the 802.15.4 devices can select a channel that is away from the busiest 802.11 channels to reduce the interference. If the 802.15.4 devices can scan the channels to detect the interference and adaptively change their operating channel over time, the performance of the 802.15.4 transmission can be further improved.

### 5.1.2. Co-existence with Microwave Oven

Since microwave ovens operate on the 2.45 GHz frequency, we set the XBee modules to work on channel 20 with a center frequency of 2.45 GHz. To test the worst case, we put the XBee remote module right next to the microwave and set the transmit power to -10 dBm. The microwave oven has a power of 800 Watts when it is ON. The results in Table 4 show that when the microwave oven is OFF, there is no data loss (note that channel 20 also experiences less interference from the WLANs); whereas the data success ratio is reduced to 96.85% with a high standard deviation of 3.22%, when the microwave oven is ON. Therefore, when used in a smart home health care application, 802.15.4 devices should avoid channel 20 if they are close to a microwave. However, if the remote module is moved to about 2 meters away, the influence from the microwave oven can be removed. We also tested the interference of a microwave oven on the two nearby 802.15.4 channels (*i.e.*, channels 19 and 21 with center frequencies of 2.445 GHz and 2.455 GHz, respectively). The test results are listed in Table 4, showing that the data success ratio is still above 99% when channel 19 or 21 is selected. Therefore, the effect of microwave oven interference can be greatly reduced even if the 802.15.4 devices choose a channel that is close to 2.45 GHz.

Table 4   Microwave oven interference test

| 802.15.4 channel (power level = - 10 dBm) | Data success ratio | | | |
|---|---|---|---|---|
| | Microwave ON | | Microwave OFF | |
| | mean | std. | Mean | std. |
| 20 | 96.85% | 3.22% | 100% | 0% |
| 19 | 99.51% | 0.31% | 100% | 0% |
| 21 | 99.34% | 0.19% | 100% | 0% |

## 5.2. Signal Attenuation Tests

A wireless body sensor system needs to operate at a low power for safety and security reasons. On the other hand, the wireless signal can be weakened by distance and the properties of the medium through which the signal travels. To evaluate the use of the 802.15.4 wireless sensor system in a home environment, we tested the signal attenuation in another single house, whose interior walls and ceilings are built with wood materials, whereas the exterior walls have aluminium siding except that one side has bricks. The 802.15.4 devices were operated at -10 dBm power level unless otherwise specified.

We first put the remote module outside the house whereas the base module is inside on the first floor. We found that when the remote module is behind a wall with aluminium siding, the signal is attenuated significantly to the point that the data success ratio is close to 0%. However, if the two modules are only separated by a brick wall or a glass window, the data success ratio is only affected slightly (still above 99.5%). When the remote module is inside the house, the data success ratio is high (above 99%) even if the signal has to pass through several wooden walls. In summary, our test results verify prior studies that building materials attenuate electromagnetic signals differently. For example, according to the attenuation test performed at the U.S. National Institute of Standards and Technology (NIST)'s Gaithersburg laboratories [24], the signal attenuation for 13 mm drywall and plywood is below 1 dB around 2 GHz frequency band; for 13 mm glass and brick walls the signal attenuation can be greater than 3 dB and 5 dB, respectively; but for concrete and reinforced concrete walls with steel ReBar mesh, the attenuation can easily exceed 30 dB. Therefore, if a house is built using steel frames or with a concrete structure (e.g., high-rise apartment buildings and hurricane-resistant houses), the 802.15.4 transmission performance is expected to experience greater degradation by the house structure. In such cases, the utilization of IEEE 802.15.4 for smart home medical applications needs to be re-evaluated.

In addition, we found two additional sources of attenuation that greatly impact the signal transmission. First, when the remote module is put close to the stove in the kitchen and the signal has to pass through the metal surface, the data success ratio drops below 75%. This is because metal obstacles tend to reflect and absorb the electromagnetic signal. The second source of attenuation found was a plant (Peace Lily) inside the house. This plant has large leaves and is about 1 meter high with the pot. When the remote module is located on a chair behind the plant as shown in Fig. 9, the data success ratio drops significantly even if the transmit power is increased to 0 dBm. However, the affecting range is short: The signal attenuation is very small if the remote module was moved to 0.5 meters away. One explanation for such result is that the water content inside the leaves absorbs and scatters the wireless signal [25][26]. Therefore, when wireless body sensor network systems are deployed for smart home usage, one suggestion to compensate the signal attenuation is placing repeaters/routers at strategic locations (*e.g*., up on a wall or on the ceiling) to avoid passing through metal appliances and big plants and to increase the transmission range.

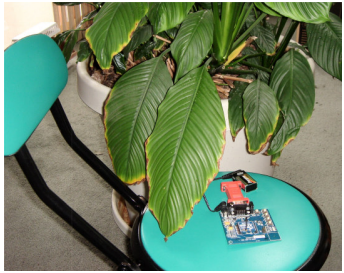

Figure 9   The plant in the house and the location of the remote module

## 6. CONCLUSIONS AND FUTURE WORK

This paper describes the design and the testing of a wireless body sensor system to acquire data concerning the physical activities of a person in need of medical care at home. This system is inexpensive, energy efficient, and unobtrusive to the human subject. The experimental tests verify that acceleration measurements can be used to classify human movements. In addition, when IEEE 802.15.4 devices are used at home, there is wireless interference from nearby IEEE 802.11 signals and microwave ovens. The data delivery performance, however, is satisfactory and can be improved by selecting an appropriate channel. We also find that housing materials, home appliances, and even plants can attenuate the wireless signal at different scales. Therefore, the deployment of wireless body sensor network systems at home needs to take these factors into account.

For future work, we plan to investigate power saving mechanisms (*e.g.*, pre-processing data in the microcontroller to reduce the amount of data to be transmitted over the air). We will study advanced numerical and statistical analysis tools to characterize and classify different activities and detect abnormal activities. We will also explore in greater detail the interference and signal attenuation issues of wireless sensor networks for smart home health care applications.